\documentclass[aps,prl,twocolumn,groupedaddress,showpacs,amsmath,amssymb,floatfix,superscriptaddress]{revtex4}
\usepackage{graphicx}
\usepackage{times}

\newcommand{\nuebar}{$\overline{\nu}_{e}$}
\newcommand{\DeltaMSq}{$\Delta m^{2}$}
\newcommand{\LiHe}{$^9$Li/$^8$He\ }
\newcommand{\BN}{$^{12}$B/$^{12}$N\ }

\newcommand{\ThetaParam}{$\tan^2 \theta$}

\newcommand{\PearsonChiSq}{$\chi_p^2$}

\begin{document}

\title{Measurement of Neutrino Oscillation with KamLAND:\\
  Evidence of Spectral Distortion}

\newcommand{\tohoku}{\affiliation{Research Center for Neutrino
    Science, Tohoku University, Sendai 980-8578, Japan}}
\newcommand{\alabama}{\affiliation{Department of Physics and
    Astronomy, University of Alabama, Tuscaloosa, Alabama 35487, USA}}
\newcommand{\lbl}{\affiliation{Physics Department, University of
    California at Berkeley and \\ Lawrence Berkeley National Laboratory, Berkeley, California 94720, USA}}
\newcommand{\caltech}{\affiliation{W.~K.~Kellogg Radiation Laboratory,
    California Institute of Technology, Pasadena, California 91125, USA}}
\newcommand{\drexel}{\affiliation{Physics Department, Drexel
    University, Philadelphia, Pennsylvania 19104, USA}}
\newcommand{\hawaii}{\affiliation{Department of Physics and Astronomy,
    University of Hawaii at Manoa, Honolulu, Hawaii 96822, USA}}
\newcommand{\lsu}{\affiliation{Department of Physics and Astronomy,
    Louisiana State University, Baton Rouge, Louisiana 70803, USA}}
\newcommand{\unm}{\affiliation{Physics Department, University of New
    Mexico, Albuquerque, New Mexico 87131, USA}}
\newcommand{\stanford}{\affiliation{Physics Department, Stanford
    University, Stanford, California 94305, USA}}
\newcommand{\ut}{\affiliation{Department of Physics and
    Astronomy, University of Tennessee, Knoxville, Tennessee 37996, USA}}
\newcommand{\tunl}{\affiliation{Triangle Universities Nuclear
    Laboratory, Durham, North Carolina 27708, USA and \\
Physics Departments at Duke University, North Carolina State University,
and the University of North Carolina at Chapel Hill}}
\newcommand{\ihep}{\affiliation{Institute of High Energy Physics,
    Beijing 100039, People's Republic of China}}
\newcommand{\cnrs}{\affiliation{CEN Bordeaux-Gradignan, IN2P3-CNRS and
    University Bordeaux I, F-33175 Gradignan Cedex, France}}

\newcommand{\aticrrnow}{\altaffiliation{Present address: ICRR, 
    University of Tokyo, Gifu, Japan}}
\newcommand{\aticeppnow}{\altaffiliation{Present address: ICEPP,
    University of Tokyo, Tokyo, Japan}}
\newcommand{\atimperialnow}{\altaffiliation{Present address: Imperial
    College London, UK}}
\newcommand{\atlanlnow}{\altaffiliation{Present address: LANL, Los
    Alamos, NM 87545, USA}}
\newcommand{\atiasnow}{\altaffiliation{Present address: School of
    Natural Sciences, Institute for Advanced Study, Princeton, NJ
    08540, USA}}
\newcommand{\atksunow}{\altaffiliation{Present address: KSU, Manhattan, KS 66506, USA}}

\author{T.~Araki}\tohoku
\author{K.~Eguchi}\tohoku
\author{S.~Enomoto}\tohoku
\author{K.~Furuno}\tohoku
\author{K.~Ichimura}\tohoku
\author{H.~Ikeda}\tohoku
\author{K.~Inoue}\tohoku
\author{K.~Ishihara}\aticrrnow\tohoku
\author{T.~Iwamoto}\aticeppnow\tohoku
\author{T.~Kawashima}\tohoku
\author{Y.~Kishimoto}\tohoku
\author{M.~Koga}\tohoku
\author{Y.~Koseki}\tohoku
\author{T.~Maeda}\tohoku
\author{T.~Mitsui}\tohoku
\author{M.~Motoki}\tohoku
\author{K.~Nakajima}\tohoku
\author{H.~Ogawa}\tohoku
\author{K.~Owada}\tohoku
\author{J.-S.~Ricol}\tohoku
\author{I.~Shimizu}\tohoku
\author{J.~Shirai}\tohoku
\author{F.~Suekane}\tohoku
\author{A.~Suzuki}\tohoku
\author{K.~Tada}\tohoku
\author{O.~Tajima}\tohoku
\author{K.~Tamae}\tohoku
\author{Y.~Tsuda}\tohoku
\author{H.~Watanabe}\tohoku
\author{J.~Busenitz}\alabama
\author{T.~Classen}\alabama
\author{Z.~Djurcic}\alabama
\author{G.~Keefer}\alabama
\author{K.~McKinny}\alabama
\author{D.-M.~Mei}\atlanlnow\alabama
\author{A.~Piepke}\alabama
\author{E.~Yakushev}\alabama
\author{B.E.~Berger}\lbl
\author{Y.D.~Chan}\lbl
\author{M.P.~Decowski}\lbl
\author{D.A.~Dwyer}\lbl
\author{S.J.~Freedman}\lbl
\author{Y.~Fu}\lbl
\author{B.K.~Fujikawa}\lbl
\author{J.~Goldman}\lbl
\author{F.~Gray}\lbl
\author{K.M.~Heeger}\lbl
\author{K.T.~Lesko}\lbl
\author{K.-B.~Luk}\lbl
\author{H.~Murayama}\atiasnow\lbl
\author{A.W.P.~Poon}\lbl
\author{H.M.~Steiner}\lbl
\author{L.A.~Winslow}\lbl
\author{G.A.~Horton-Smith}\atksunow\caltech
\author{C.~Mauger}\caltech
\author{R.D.~McKeown}\caltech
\author{P.~Vogel}\caltech
\author{C.E.~Lane}\drexel
\author{T.~Miletic}\drexel
\author{P.W.~Gorham}\hawaii
\author{G.~Guillian}\hawaii
\author{J.G.~Learned}\hawaii
\author{J.~Maricic}\hawaii
\author{S.~Matsuno}\hawaii
\author{S.~Pakvasa}\hawaii
\author{S.~Dazeley}\lsu
\author{S.~Hatakeyama}\lsu
\author{A.~Rojas}\lsu
\author{R.~Svoboda}\lsu
\author{B.D.~Dieterle}\unm
\author{J.~Detwiler}\stanford
\author{G.~Gratta}\stanford
\author{K.~Ishii}\stanford
\author{N.~Tolich}\stanford
\author{Y.~Uchida}\atimperialnow\stanford
\author{M.~Batygov}\ut
\author{W.~Bugg}\ut
\author{Y.~Efremenko}\ut
\author{Y.~Kamyshkov}\ut
\author{A.~Kozlov}\ut
\author{Y.~Nakamura}\ut
\author{C.R.~Gould}\tunl
\author{H.J.~Karwowski}\tunl
\author{D.M.~Markoff}\tunl
\author{J.A.~Messimore}\tunl
\author{K.~Nakamura}\tunl
\author{R.M.~Rohm}\tunl
\author{W.~Tornow}\tunl
\author{R.~Wendell}\tunl
\author{A.R.~Young}\tunl
\author{M.-J.~Chen}\ihep
\author{Y.-F.~Wang}\ihep
\author{F.~Piquemal}\cnrs

\collaboration{The KamLAND Collaboration}\noaffiliation

\date{\today}

\begin{abstract}

We present results of a study of neutrino oscillation based on a
766 ton-year exposure of KamLAND to reactor
anti-neutrinos. We observe 258 \nuebar\ candidate events
with energies above 3.4\,MeV compared to 365.2 events
expected in the absence of neutrino oscillation. Accounting for 17.8
expected background events, the statistical significance for reactor
\nuebar\ disappearance is 99.998\%. The observed energy
spectrum disagrees with the expected spectral shape in the absence of
neutrino oscillation at 99.6\% significance and prefers
the distortion expected from \nuebar\ oscillation effects. A
two-neutrino oscillation analysis of the KamLAND data gives
\DeltaMSq\,=\, 7.9$^{+0.6}_{-0.5}\times$10$^{-5}$\,eV$^2$. A global analysis of data from
KamLAND and solar neutrino experiments yields
\DeltaMSq\,=\,7.9$^{+0.6}_{-0.5}\times$10$^{-5}$\,eV$^2$ and
\ThetaParam\,=\,0.40$^{+0.10}_{-0.07}$, the most precise determination to
date.

\end{abstract}


\maketitle

The first measurement of reactor anti-neutrino disappearance by
KamLAND~\cite{Eguchi} suggested that solar neutrino flavor
transformation through the Mikheyev-Smirnov-Wolfenstein
(MSW)~\cite{MSW} matter effect has a direct correspondence to
anti-neutrino oscillation in vacuum. Assuming CPT invariance, KamLAND
and solar-neutrino experiments have restricted the solar oscillation
parameters, eliminating all but the large-mixing-angle (LMA-MSW)
solution. This Letter reports more stringent constraints on neutrino
oscillation parameters from KamLAND based on a three times longer
exposure and a 33\% increase in the usable fiducial
volume. Large variations in the reactor power production in Japan in
2003 allowed us to study the \nuebar\ flux dependence. The first
evidence for spectral distortion in the \nuebar\ spectrum is provided
here; spectral distortion is direct evidence of an oscillation effect.

KamLAND consists of 1\,kton of ultra-pure liquid scintillator (LS)
contained in a 13-m-diameter transparent nylon-based balloon suspended
in non-scintillating oil. The balloon is surrounded by 1879
photomultiplier tubes (PMTs) mounted on the inner surface of an
18-m-diameter spherical stainless-steel vessel.  Electron
anti-neutrinos are detected via inverse $\beta$-decay,
$\overline{\nu}_e + p\rightarrow e^+ + n$, with a 1.8\,MeV
\nuebar\ energy threshold. The prompt scintillation light from the
$e^+$ gives an estimate of the incident \nuebar\ energy,
$E_{\overline{\nu}_e} = E_{\text{prompt}} + \overline{E}_n +
0.8$\,MeV, where $E_{\text{prompt}}$ is the prompt event energy
including the positron kinetic energy and the annihilation energy, and
$\overline{E}_n$ is the average neutron recoil energy. The
$\sim200\,\mu s$ delayed 2.2\,MeV $\gamma$ ray from neutron capture on
hydrogen is a powerful tool for reducing background. On average,
neutrons are captured within 9\,cm and the spatial
correlation between prompt and delayed signals is dominated by the
vertex position resolution. A 3.2\,kton water-Cherenkov detector
surrounds the containment sphere, absorbing $\gamma$ rays and neutrons
from the surrounding rock and tagging cosmic-ray muons.  This outer
detector (OD) is over 92\% efficient for muons
passing through the fiducial volume.

KamLAND is surrounded by 53 Japanese power reactor units. The reactor
operation data, including thermal power generation, fuel burn up,
exchange and enrichment records, are provided by all Japanese
power reactors and are used to calculate fission rates of
each isotope. The averaged relative fission yields for the run period
were $^{235}$U\,:\,$^{238}$U\,:\,$^{239}$Pu\,:\,$^{241}$Pu =
0.563\,:\,0.079\,:\,0.301\,:\,0.057. The expected \nuebar\ flux is
calculated using fission rates and \nuebar\ spectra; the spectra are
from Ref.~\cite{spectra}. The \nuebar\ contribution
from Japanese research reactors and all reactors outside of Japan is
4.5\%. We assume that these reactors have the same average fuel
composition as the Japanese power reactors. The total integrated
thermal power flux over the detector livetime was
701\,Joule/cm$^{2}$.

We report on data collected between March 9, 2002 and January 11,
2004, including re-analysis of the data used in Ref.~\cite{Eguchi}. The
central detector PMT array was upgraded on February 27, 2003 by
commissioning 554 20-inch PMTs, increasing the photo-cathode coverage
from 22\% to 34\% and improving the energy resolution from
7.3\%$/\sqrt{E(\text{MeV})}$ to
6.2\%$/\sqrt{E(\text{MeV})}$. The trigger threshold of
200 hit 17-inch PMTs corresponds to about 0.7\,MeV at the detector
center and has an efficiency close to 100\%. We use a prompt event
energy analysis threshold of 2.6\,MeV to avoid backgrounds including
the effect of anti-neutrinos from uranium and thorium decaying in the
Earth (geo-neutrinos).

The location of interactions inside the detector is determined from
PMT hit timing; the energy is obtained from the number of observed
photo-electrons after correcting for position and gain
variations. Position and time dependence of the energy estimation are
monitored periodically with $\gamma$-ray and neutron sources along the
central vertical axis (z-axis) of the scintillator volume. Trace
radio-isotopes on the balloon and in the scintillator are also
exploited. The systematic uncertainty in the energy scale at the
2.6\,MeV prompt event energy
($E_{\overline{\nu}_e}\,\simeq$\,3.4\,MeV) analysis threshold is
2.0\%, corresponding to a 2.3\% uncertainty
in the number of events in an unoscillated reactor \nuebar\ spectrum.

The radial fiducial volume cut is relaxed from 5\,m \cite{Eguchi} to
5.5\,m in the present analysis, expanding the fiducial mass to
543.7\,tons (4.61$\times$10$^{31}$ free target protons). The
radial positions of the prompt and delayed event are both required to
be less than 5.5\,m.  The 1.2\,m cylindrical cut along the z-axis
previously used to exclude low energy backgrounds from thermometers is
not applied. The event selection cuts for the time difference ($\Delta
T$) and position difference ($\Delta R$) between the positron and
delayed neutron are 0.5\,$\mu$s\,$<\Delta T<$\,1000\,$\mu$s and
$\Delta R\,<\,$2\,m, respectively. The event energies are required to
be 1.8\,MeV$\,<\,E_{\text{delayed}}\,<\,$2.6\,MeV and
2.6\,MeV$\,<\,E_{\text{prompt}}\,<$\,8.5\,MeV. The efficiency of all
cuts is (89.8$\,\pm$\,1.5)\%.

The total volume of the liquid scintillator is
1171\,$\pm$\,25\,m$^3$, as measured by flow meters during detector
filling.  The nominal 5.5-m-radius fiducial volume
($\frac{4}{3}\pi R^{3}$) corresponds to 0.595$\,\pm\,$0.013 of the
total LS volume.  The effective fiducial volume is defined by the cuts on
the radial positions of the reconstructed event vertices. At present,
only z-axis calibrations are available, so we assess the systematic
uncertainty in the total fiducial volume by studying
uniformly-distributed muon spallation products, identified as delayed
coincidences following muons.  We measure the position
distribution of the $\beta$-decays of $^{12}$B ($Q$\,=\,13.4\,MeV,
$\tau_{1/2}$\,=\,20.2\,ms) and $^{12}$N ($Q$\,=\,17.3\,MeV,
$\tau_{1/2}$\,=\,11.0\,ms), which are produced at
the rate of about 60~\BN\ events/kton-day.  Fits to the energy
distribution of these events indicate that the sample is mostly
$^{12}$B; the relative contribution of $^{12}$N is only
$\sim$1\%.  The number of \BN\ events reconstructed in the
fiducial volume compared to the total number in the entire LS volume
is 0.607\,$\pm$\,0.006(stat)\,$\pm$\,0.006(syst), where the systematic
error arises from events near the balloon surface that deposit a fraction
of their energy outside the LS.  As a consistency check, in a similar
study of spallation neutrons, which we identify via the 2.2\,MeV
capture $\gamma$ ray, we find the ratio 0.587\,$\pm$\,0.013(stat).

The \BN\ events typically have higher energy than \nuebar\
candidates, so an additional systematic error accounts
for possible dependence of effective fiducial volume on energy.  We
constrain the variation to 2.7\% by comparing the prompt and delayed
event positions of delayed-neutron $\beta$-decays of $^9$Li
($Q$\,=\,13.6\,MeV, $\tau_{1/2}$\,=\,178\,ms) and $^8$He
($Q$\,=\,10.7\,MeV, $\tau_{1/2}$\,=\,119\,ms). The observed capture
distance variation is a measure of the energy uniformity of the vertex
finding algorithm. Combining the errors from the LS volume
measurements, the \BN\ volume ratio calibration, and the constraints
on energy dependence, we obtain a 4.7\% systematic error
on the fiducial volume.

The rate of accidental coincidences increases in the outer region of
the fiducial volume, since most background sources are external to the
liquid scintillator. This background is estimated with a 10\,ms to
20\,s delayed-coincidence window and by pairing random singles
events. These consistent methods predict 2.69\,$\pm$\,0.02 events above
the 2.6\,MeV threshold.

Above 2.6\,MeV, neutrons and long-lived delayed-neutron
$\beta$-emitters are sources of correlated backgrounds. The
$\sim$3000 spallation-produced neutrons per kton-day are effectively
eliminated with a 2\,ms veto of the entire detector following a
detected muon. The remaining fast neutrons come from muons
missed by the OD or interacting in the rock just outside it. This
background is reduced significantly by the OD
and several layers of absorbers: the OD itself, the 2.5\,m of
non-scintillating oil surrounding the LS, and the 1\,m of LS outside
the fiducial volume. We estimate this background contributes fewer
than 0.89 events to the data sample.

The uncorrelated background from \BN\ spallation products is
effectively suppressed by the delayed-coincidence
requirement. However, the $\sim$1.5 events/kton-day in the
delayed-neutron branches of $^{9}$Li and $^{8}$He mimic the
\nuebar\ signal. From fits to the decay-time and $\beta$-energy
spectra we see mostly $^{9}$Li decays; the contribution of $^{8}$He
relative to $^{9}$Li is less than 15\% at 90\% C.L. For
isolated, well-tracked muons passing through the detector, we apply a
2\,s veto within a 3\,m radius cylinder around the track.  We veto the
entire volume for 2\,s after one in $\sim$30 muons, those that
produce more than $\sim$10$^{6}$ photo-electrons above minimum
ionization or muons tracked with poor reliability. We estimate that
4.8\,$\pm$\,0.9 \LiHe\ events remain after the cuts. The deadtime
introduced by all muon cuts is 9.7\%; the total livetime
including spallation cuts is 515.1\,days.

A third source of correlated background comes indirectly from the
$\alpha$ decays of the radon daughter $^{210}$Po in the liquid
scintillator. The signal of the 5.3\,MeV $\alpha$-particle is quenched
below the threshold, but the secondary reaction
$^{13}$C($\alpha$,n)$^{16}$O produces events above 2.6\,MeV. Special
runs to observe the decay of $^{210}$Po establish that there were
(1.47$\pm$0.20)$\times$10$^{9}$ $\alpha$ decays during the livetime of data
taking. Using the $^{13}$C($\alpha$,n) reaction cross sections from
Ref.~\cite{Sekharan}, Monte Carlo simulations and detailed studies of
quenching effects to convert the outgoing neutron energy spectrum into
a visible energy spectrum, we expect 10.3\,$\pm$\,7.1 events above
2.6\,MeV. The spectrum is concentrated around 6\,MeV, from decays of
levels in $^{16}$O, and 4.4\,MeV, from $\gamma$ decays following
neutron inelastic scattering on $^{12}$C. The observed energy from
neutron-proton elastic scattering is mostly quenched below
2.6\,MeV. This $\alpha$-induced background was not considered in
Ref.~\cite{Eguchi} and would have contributed 1.9\,$\pm$\,1.3
additional background events (2.8\,$\pm$\,1.7 total background
events). The total background to the \nuebar-signal above 2.6\,MeV in
the present analysis is 17.8\,$\pm$\,7.3 events, where the bound on
the fast neutron background is accounted for in the uncertainty.

\begin{table}[tb]
\caption{Estimated systematic uncertainties (\%). }
\label{tab:sys}
\begin{tabular}{llll}
\hline \hline
Fiducial Volume           &  4.7   &\ \ \ \ Reactor power      &  2.1    \\
Energy threshold          & 2.3  &\ \ \ \ Fuel composition                    &  1.0    \\
Efficiency of cuts        & 1.6  &\ \ \ \ $\overline{\nu}_e$ spectra \cite{spectra}   &  2.5    \\
Livetime      & 0.06  &\ \ \ \ Cross section~\cite{crosssection}        &  0.2    \\
\hline
Total systematic uncertainty  &        &                                            & 6.5 \\
\hline \hline
\end{tabular}
\end{table}

In the absence of anti-neutrino disappearance, we expect to observe
365.2\,$\pm$\,23.7(syst) \nuebar\ events above 2.6\,MeV, where the systematic
uncertainty is detailed in Table~\ref{tab:sys}. We observe
258 events, confirming \nuebar\ disappearance at the
99.998\% significance level. The average \nuebar\ survival
probability is 0.658\,$\pm$\,0.044(stat)\,$\pm$\,0.047(syst), where the background error has been
included in the systematic uncertainty. The effective baseline varies
with power output of the reactor sources involved, so the survival
probabilities for different periods are not directly
comparable. Applying the new analysis on the previously reported data
(March 2002 to October 2002) \cite{Eguchi} gives 0.601\,$\pm$\,0.069(stat)\,$\pm$\,0.042(syst), in
agreement with 0.589\,$\pm$\,0.085(stat)\,$\pm$\,0.042(syst), after correction for the ($\alpha$,n)
background.

\begin{figure}
\includegraphics[width=8cm]{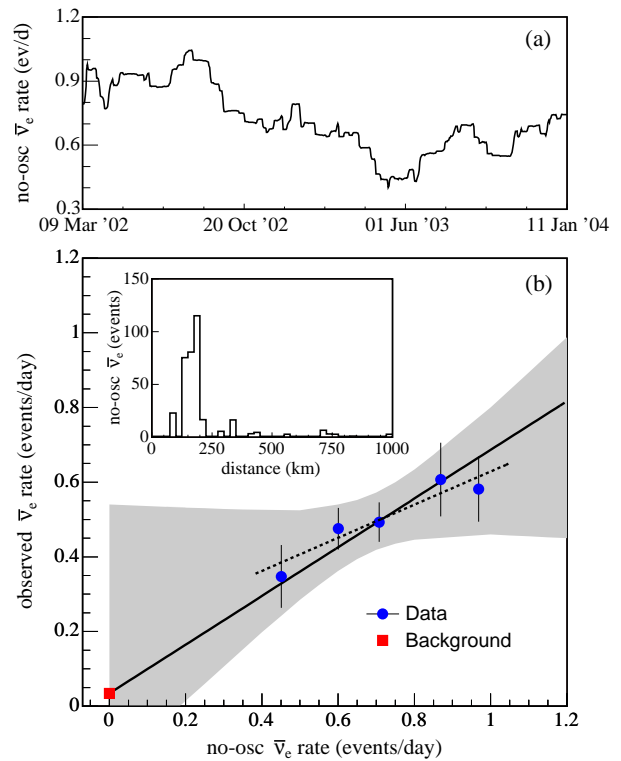}
\put(-33,270){(a)}
\put(-33,177){(b)}
\caption[] {(a) Estimated time variation of the reactor \nuebar\ flux
  at KamLAND assuming no anti-neutrino oscillation. (b) Observed
  \nuebar\ event rate versus no-oscillation reactor
  \nuebar\ flux. Data points correspond to intervals of approximately
  equal \nuebar\ flux. The dashed line is a fit, the
  90\% C.L. is shown in gray. The solid line is a fit constrained to
  the expected background. The reactor
  distance distribution for \nuebar\ events in the absence of
  oscillations is shown in the inset.}
\label{fig:timevar}
\end{figure}

After September 2002, a number of Japanese nuclear reactors were
off, as indicated in Fig.~\ref{fig:timevar}a. This decreased the
expected no-oscillation \nuebar\ flux by more than a factor of two. In
Fig.~\ref{fig:timevar}b the signal counts are plotted in bins of
approximately equal \nuebar\ flux corresponding to total reactor
power. For \DeltaMSq\ and \ThetaParam\ determined below and the known
distributions of reactor power level and distance, the expected
oscillated \nuebar\ rate is well approximated by a straight line. The
slope can be interpreted as the reactor-correlated signal and the
intercept as the reactor-independent constant background
rate. Fig.~\ref{fig:timevar}b shows the linear fit and its 90\%
C.L. region. The intercept is consistent with known backgrounds, but
substantially larger backgrounds cannot be excluded; hence this fit
does not usefully constrain speculative sources of anti-neutrinos such
as a nuclear reactor at the Earth's core~\cite{georeactor}. The
predicted KamLAND rate for typical 3\,TW geo-reactor scenarios is
comparable to the expected 17.8\,$\pm$\,7.3 event background and would
have minimal impact on the analysis of the reactor power dependence
signal. In the following we consider contributions only from known
anti-neutrino sources.

\begin{figure}
\includegraphics[width=8cm]{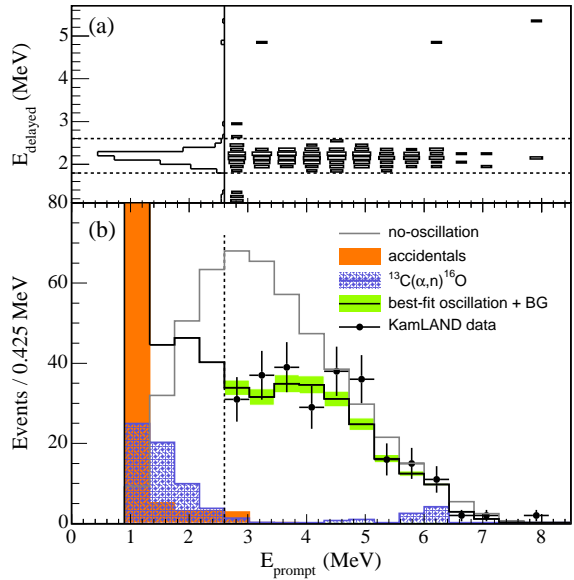}
\put(-194,207){(a)}
\put(-194,130){(b)}
\caption[]{(a) The correlation between the prompt and delayed event
  energies after cuts. The three events with
  $E_{\text{delayed}}\sim5$\,MeV are consistent with neutron capture on
  carbon. (b) Prompt event energy spectrum of \nuebar\ candidate
  events with associated background spectra. The shaded band indicates
  the systematic error in the best-fit reactor spectrum above
  2.6\,MeV.}
\label{fig:spectrum}
\end{figure}

Fig.~\ref{fig:spectrum}a shows the correlation of the prompt and
delayed event energy after all selection cuts except for the
$E_{\text{delayed}}$ cut. The prompt energy spectrum above 2.6 MeV is
shown in Fig.~\ref{fig:spectrum}b. The data evaluation method with an
unbinned maximum likelihood fit to two-flavor neutrino oscillation is
similar to the method used previously~\cite{Eguchi}. In the present
analysis, we account for the $^{9}$Li, accidental and the
$^{13}$C($\alpha$,n)$^{16}$O background rates. For the ($\alpha$,n)
background, the contribution around 6\,MeV is allowed to float because
of uncertainty in the cross section, while the contributions around
2.6\,MeV and 4.4\,MeV are constrained to within 32\% of the estimated
rate. We allow for a 10\% energy scale uncertainty for the 2.6\,MeV
contribution due to neutron quenching uncertainty. The best-fit
spectrum together with the backgrounds is shown in
Fig.~\ref{fig:spectrum}b; the best-fit for the rate-and-shape analysis
is \DeltaMSq\,=\,7.9$^{+0.6}_{-0.5}\times$10$^{-5}$\,eV$^{2}$ and
\ThetaParam\,=\,0.46, with a large uncertainty on
\ThetaParam. A shape-only analysis gives
\DeltaMSq\,=\,(8.0\,$\pm$\,0.5)$\times$10$^{-5}$\,eV$^2$ and
\ThetaParam\,=\,0.76.

Taking account of the backgrounds, the Baker-Cousins $\chi^2$ for the
best-fit is 13.1 (11 DOF). To test the goodness-of-fit we follow
the statistical techniques in Ref.~\cite{Kendall}. First, the data are fit
to a hypothesis to find the best-fit parameters.  Next, we bin the
energy spectrum of the data into 20 equal-probability bins and
calculate the Pearson $\chi^2$ statistic (\PearsonChiSq) for the
data. Based on the particular hypothesis 10,000 spectra were generated
using the parameters obtained from the data and \PearsonChiSq\ was
determined for each spectrum.  The confidence level of the data is the
fraction of simulated spectra with a higher \PearsonChiSq.  For the
best-fit oscillation parameters and the {\it a priori} choice of 20 bins,
the goodness-of-fit is 11.1\% with
$\chi_\text{p}^2$/DOF\,=\,24.2/17.  The goodness-of-fit of
the scaled no-oscillation spectrum where the normalization was fit to
the data is 0.4\%
($\chi_\text{p}^2$/DOF\,=\,37.3/18). We note that
the \PearsonChiSq\ and goodness-of-fit results are sensitive to the
choice of binning.

\begin{figure}
\includegraphics[width=8cm]{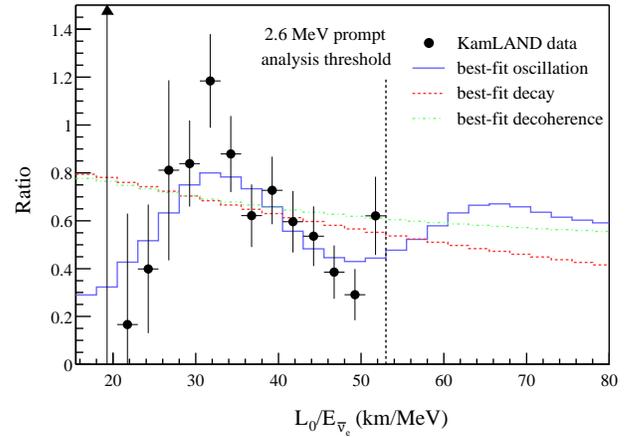}
\caption[] {Ratio of the observed \nuebar\ spectrum to the
expectation for no-oscillation versus L$_{0}$/E. The curves show the
expectation for the best-fit oscillation, best-fit decay and best-fit
decoherence models taking into account the individual time-dependent
flux variations of all reactors and detector effects. The data points
and models are plotted with L$_{0}$=180\,km, as if all anti-neutrinos
detected in KamLAND were due to a single reactor at this distance.}
\label{fig:LE}
\end{figure}

\begin{figure*}
\includegraphics[width=16cm]{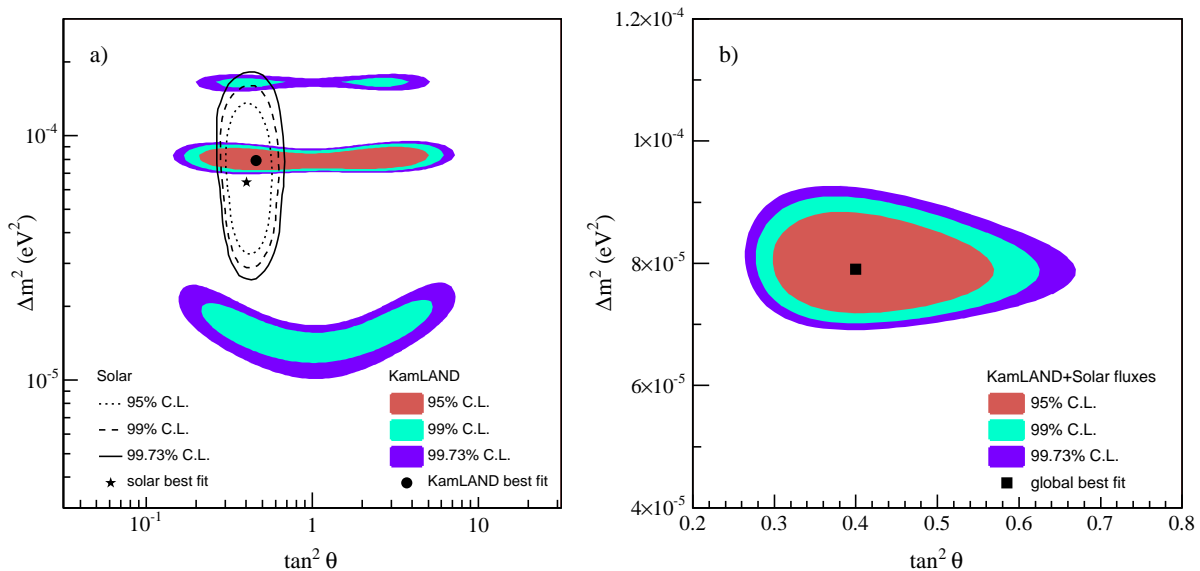}
\put(-423,195){a)}
\put(-185,195){b)}
\caption{(a) Neutrino oscillation parameter allowed region from
  KamLAND anti-neutrino data (shaded regions) and solar neutrino
  experiments (lines) \cite{snosalt}. (b) Result of a combined
  two-neutrino oscillation analysis of KamLAND and the observed solar
  neutrino fluxes under the assumption of CPT invariance. The fit
  gives \DeltaMSq\,=\,7.9$^{+0.6}_{-0.5}\times$10$^{-5}$\,eV$^{2}$ and
  \ThetaParam\,=\,0.40$^{+0.10}_{-0.07}$ including the allowed 1-sigma
  parameter range.}
\label{fig:3Nu}
\end{figure*}

To illustrate oscillatory behavior of the data, we plot in
Fig.~\ref{fig:LE} the L$_{0}$/E distribution, where the data and the
best-fit spectra are divided by the expected no-oscillation
spectrum. Two alternative hypotheses for neutrino disappearance,
neutrino decay~\cite{NeutrinoDecay} and
decoherence~\cite{NeutrinoDecoherence}, give different L$_{0}$/E
dependences.  As in the oscillation analysis, we survey the parameter
spaces and find the best-fit points at $(\sin^2\theta,\,m/c\tau)$ =
(1.0,\,0.011\,MeV/km) for decay and $(\sin^22\theta,\,\gamma^0)$ =
(1.0,\,0.030\,MeV/km) for decoherence, using the notation of the
references.  Applying the goodness-of-fit procedure described above,
we find that decay has a goodness-of-fit of only 0.7\%
($\chi_\text{p}^2$/DOF\,=\,35.8/17), while decoherence has a
goodness-of-fit of 1.8\%
($\chi_\text{p}^2$/DOF\,=\,32.2/17). We note that, while the
present best-fit neutrino decay point has already been ruled out by
solar neutrino data~\cite{NuDecayBeacom} and observation of SN1987A,
the decay model is used here as an example of a scenario resulting in
a \nuebar\ deficit.  If we do not assume CPT invariance and allow the
range 0.5\,$<\,\sin^2\theta\,<$\,0.75, then the decay scenario
considered here can avoid conflict with solar
neutrino~\cite{NuDecayBeacom} and SN1987A data~\cite{NuDecayCPT}.

The allowed region contours in \DeltaMSq-\ThetaParam\ parameter space
derived from the $\Delta \chi^2$ values (e.g., $\Delta
\chi^2\,<\,$5.99 for 95\% C.L.) are shown in Fig.~\ref{fig:3Nu}a.  The
best-fit point is in the region commonly characterized as
LMA~I. Maximal mixing for values of \DeltaMSq\ consistent with LMA~I
is allowed at the 62.1\% C.L.  Due to distortions in the
spectrum, the LMA~II region (at
\DeltaMSq$\sim$2$\times$10$^{-4}$\,eV$^{2}$) is disfavored at the
98.0\% C.L., as are larger values of \DeltaMSq\ previously
allowed by KamLAND.  The allowed region at lower \DeltaMSq\ is
disfavored at the 97.5\% C.L., but this region is not
consistent with the LMA region determined from solar neutrino
experiments assuming CPT invariance.

A two-flavor analysis of the KamLAND data and the observed solar
neutrino fluxes~\cite{GlobalAnalysis}, with the assumption of CPT
invariance, restricts the allowed \DeltaMSq-\ThetaParam\ parameters in
Fig. 4b. The sensitivity in \DeltaMSq\ is dominated by the observed
distortion in the KamLAND spectrum, while solar neutrino data provide
the best constraint on $\theta$. The combined analysis gives
\DeltaMSq\,=\,7.9$^{+0.6}_{-0.5}\times$10$^{-5}$\,eV$^2$ and
\ThetaParam\,=\,0.40$^{+0.10}_{-0.07}$.

The conclusion that the LMA~II region is excluded is strengthened by
the present result. The observed distortion of the spectral shape
supports the conclusion that the observation of reactor \nuebar\
disappearance is due to neutrino oscillation. Statistical
uncertainties in the KamLAND data are now on the same level as
systematic uncertainties.  Current efforts to perform full-volume source
calibrations and a reevaluation of reactor power uncertainties should
reduce the systematic uncertainties.

The KamLAND experiment is supported by the COE program under grant
09CE2003 of the Japanese Ministry of Education, Culture, Sports,
Science and Technology, and under the United States Department of
Energy grant DEFG03-00ER41138.  The reactor data are provided by
courtesy of the following electric associations in Japan: Hokkaido,
Tohoku, Tokyo, Hokuriku, Chubu, Kansai, Chugoku, Shikoku and Kyushu
Electric Power Companies, Japan Atomic Power Co. and Japan Nuclear
Cycle Development Institute. The Kamioka Mining and Smelting Company has
provided service for activities in the mine.

\end{document}